\documentclass[10pt,conference,twocolumn]{IEEEtran}

\usepackage{cite}
\usepackage[T1]{fontenc}
\usepackage{graphicx}
\usepackage{amssymb}
\usepackage{amsmath}
\usepackage{amsthm}
 
\usepackage{microtype}
\usepackage{balance}
\usepackage{subfigure}
\usepackage{xcolor}
\usepackage{booktabs}
\usepackage{soul}
\usepackage{algorithm}

\usepackage{amssymb}
\usepackage{amsfonts}
\usepackage{mathrsfs}
\usepackage{xspace}
\usepackage{bm}
\usepackage{upgreek}

\newcommand{\safemath}[2]{\newcommand{#1}{\ensuremath{#2}\xspace}}

\safemath{\bma}{\mathbf{a}}
\safemath{\bmb}{\mathbf{b}}
\safemath{\bmc}{\mathbf{c}}
\safemath{\bmd}{\mathbf{d}}
\safemath{\bme}{\mathbf{e}}
\safemath{\bmf}{\mathbf{f}}
\safemath{\bmg}{\mathbf{g}}
\safemath{\bmh}{\mathbf{h}}
\safemath{\bmi}{\mathbf{i}}
\safemath{\bmj}{\mathbf{j}}
\safemath{\bmk}{\mathbf{k}}
\safemath{\bml}{\mathbf{l}}
\safemath{\bmm}{\mathbf{m}}
\safemath{\bmn}{\mathbf{n}}
\safemath{\bmo}{\mathbf{o}}
\safemath{\bmp}{\mathbf{p}}
\safemath{\bmq}{\mathbf{q}}
\safemath{\bmr}{\mathbf{r}}
\safemath{\bms}{\mathbf{s}}
\safemath{\bmt}{\mathbf{t}}
\safemath{\bmu}{\mathbf{u}}
\safemath{\bmv}{\mathbf{v}}
\safemath{\bmw}{\mathbf{w}}
\safemath{\bmx}{\mathbf{x}}
\safemath{\bmy}{\mathbf{y}}
\safemath{\bmz}{\mathbf{z}}
\safemath{\bmzero}{\mathbf{0}}
\safemath{\bmone}{\mathbf{1}}

\bmdefine{\biad}{a}
\bmdefine{\bibd}{b}
\bmdefine{\bicd}{c}
\bmdefine{\bidd}{d}
\bmdefine{\bied}{e}
\bmdefine{\bifd}{f}
\bmdefine{\bigd}{g}
\bmdefine{\bihd}{h}
\bmdefine{\biid}{i}
\bmdefine{\bijd}{j}
\bmdefine{\bikd}{k}
\bmdefine{\bild}{l}
\bmdefine{\bimd}{m}
\bmdefine{\bind}{n}
\bmdefine{\biod}{o}
\bmdefine{\bipd}{p}
\bmdefine{\biqd}{q}
\bmdefine{\bird}{r}
\bmdefine{\bisd}{s}
\bmdefine{\bitd}{t}
\bmdefine{\biud}{u}
\bmdefine{\bivd}{v}
\bmdefine{\biwd}{w}
\bmdefine{\bixd}{x}
\bmdefine{\biyd}{y}
\bmdefine{\bizd}{z}

\bmdefine{\bixid}{\xi}
\bmdefine{\bilambdad}{\lambda}
\bmdefine{\bimud}{\mu}
\bmdefine{\bithetad}{\theta}
\bmdefine{\biphid}{\phi}
\bmdefine{\bideltad}{\delta}

\safemath{\bmia}{\biad}
\safemath{\bmib}{\bibd}
\safemath{\bmic}{\bicd}
\safemath{\bmid}{\bidd}
\safemath{\bmie}{\bied}
\safemath{\bmif}{\bifd}
\safemath{\bmig}{\bigd}
\safemath{\bmih}{\bihd}
\safemath{\bmii}{\biid}
\safemath{\bmij}{\bijd}
\safemath{\bmik}{\bikd}
\safemath{\bmil}{\bild}
\safemath{\bmim}{\bimd}
\safemath{\bmin}{\bind}
\safemath{\bmio}{\biod}
\safemath{\bmip}{\bipd}
\safemath{\bmiq}{\biqd}
\safemath{\bmir}{\bird}
\safemath{\bmis}{\bisd}
\safemath{\bmit}{\bitd}
\safemath{\bmiu}{\biud}
\safemath{\bmiv}{\bivd}
\safemath{\bmiw}{\biwd}
\safemath{\bmix}{\bixd}
\safemath{\bmiy}{\biyd}
\safemath{\bmiz}{\bizd}

\safemath{\bmxi}{\bixid}
\safemath{\bmlambda}{\bilambdad}
\safemath{\bmmu}{\bimud}
\safemath{\bmtheta}{\bithetad}
\safemath{\bmphi}{\biphid}
\safemath{\bmdelta}{\bideltad}

\safemath{\bA}{\mathbf{A}}
\safemath{\bB}{\mathbf{B}}
\safemath{\bC}{\mathbf{C}}
\safemath{\bD}{\mathbf{D}}
\safemath{\bE}{\mathbf{E}}
\safemath{\bF}{\mathbf{F}}
\safemath{\bG}{\mathbf{G}}
\safemath{\bH}{\mathbf{H}}
\safemath{\bI}{\mathbf{I}}
\safemath{\bJ}{\mathbf{J}}
\safemath{\bK}{\mathbf{K}}
\safemath{\bL}{\mathbf{L}}
\safemath{\bM}{\mathbf{M}}
\safemath{\bN}{\mathbf{N}}
\safemath{\bO}{\mathbf{O}}
\safemath{\bP}{\mathbf{P}}
\safemath{\bQ}{\mathbf{Q}}
\safemath{\bR}{\mathbf{R}}
\safemath{\bS}{\mathbf{S}}
\safemath{\bT}{\mathbf{T}}
\safemath{\bU}{\mathbf{U}}
\safemath{\bV}{\mathbf{V}}
\safemath{\bW}{\mathbf{W}}
\safemath{\bX}{\mathbf{X}}
\safemath{\bY}{\mathbf{Y}}
\safemath{\bZ}{\mathbf{Z}}

\safemath{\bZero}{\mathbf{0}}
\safemath{\bOne}{\mathbf{1}}
\safemath{\bDelta}{\mathbf{\Delta}}
\safemath{\bLambda}{\mathbf{\UpLambda}}
\safemath{\bPhi}{\mathbf{\Upphi}}
\safemath{\bSigma}{\mathbf{\Upsigma}}
\safemath{\bOmega}{\mathbf{\Upomega}}
\safemath{\bTheta}{\mathbf{\Uptheta}}

\bmdefine{\biAd}{A}
\bmdefine{\biBd}{B}
\bmdefine{\biCd}{C}
\bmdefine{\biDd}{D}
\bmdefine{\biEd}{E}
\bmdefine{\biFd}{F}
\bmdefine{\biGd}{G}
\bmdefine{\biHd}{H}
\bmdefine{\biId}{I}
\bmdefine{\biJd}{J}
\bmdefine{\biKd}{K}
\bmdefine{\biLd}{L}
\bmdefine{\biMd}{M}
\bmdefine{\biOd}{N}
\bmdefine{\biPd}{O}
\bmdefine{\biQd}{P}
\bmdefine{\biRd}{R}
\bmdefine{\biSd}{S}
\bmdefine{\biTd}{T}
\bmdefine{\biUd}{U}
\bmdefine{\biVd}{V}
\bmdefine{\biWd}{W}
\bmdefine{\biXd}{X}
\bmdefine{\biYd}{Y}
\bmdefine{\biZd}{Z}

\bmdefine{\biDelta}{\Delta}
\bmdefine{\biLambda}{\Lambda}
\bmdefine{\biPhi}{\Phi}
\bmdefine{\biSigma}{\Sigma}
\bmdefine{\biOmega}{\Omega}
\bmdefine{\biTheta}{\Theta}

\safemath{\bimA}{\biAd}
\safemath{\bimB}{\biBd}
\safemath{\bimC}{\biCd}
\safemath{\bimD}{\biDd}
\safemath{\bimE}{\biEd}
\safemath{\bimF}{\biFd}
\safemath{\bimG}{\biGd}
\safemath{\bimH}{\biHd}
\safemath{\bimI}{\biId}
\safemath{\bimJ}{\biJd}
\safemath{\bimK}{\biKd}
\safemath{\bimL}{\biLd}
\safemath{\bimM}{\biMd}
\safemath{\bimN}{\biNd}
\safemath{\bimO}{\biOd}
\safemath{\bimP}{\biPd}
\safemath{\bimQ}{\biQd}
\safemath{\bimR}{\biRd}
\safemath{\bimS}{\biSd}
\safemath{\bimT}{\biTd}
\safemath{\bimU}{\biUd}
\safemath{\bimV}{\biVd}
\safemath{\bimW}{\biWd}
\safemath{\bimX}{\biXd}
\safemath{\bimY}{\biYd}
\safemath{\bimZ}{\biZd}

\safemath{\bimDelta}{\biDelta}
\safemath{\bimLambda}{\biLambda}
\safemath{\bimPhi}{\biPhi}
\safemath{\bimSigma}{\biSigma}
\safemath{\bimOmega}{\biOmega}
\safemath{\bimTheta}{\biTheta}

\safemath{\setA}{\mathcal{A}}
\safemath{\setB}{\mathcal{B}}
\safemath{\setC}{\mathcal{C}}
\safemath{\setD}{\mathcal{D}}
\safemath{\setE}{\mathcal{E}}
\safemath{\setF}{\mathcal{F}}
\safemath{\setG}{\mathcal{G}}
\safemath{\setH}{\mathcal{H}}
\safemath{\setI}{\mathcal{I}}
\safemath{\setJ}{\mathcal{J}}
\safemath{\setK}{\mathcal{K}}
\safemath{\setL}{\mathcal{L}}
\safemath{\setM}{\mathcal{M}}
\safemath{\setN}{\mathcal{N}}
\safemath{\setO}{\mathcal{O}}
\safemath{\setP}{\mathcal{P}}
\safemath{\setQ}{\mathcal{Q}}
\safemath{\setR}{\mathcal{R}}
\safemath{\setS}{\mathcal{S}}
\safemath{\setT}{\mathcal{T}}
\safemath{\setU}{\mathcal{U}}
\safemath{\setV}{\mathcal{V}}
\safemath{\setW}{\mathcal{W}}
\safemath{\setX}{\mathcal{X}}
\safemath{\setY}{\mathcal{Y}}
\safemath{\setZ}{\mathcal{Z}}
\safemath{\emptySet}{\varnothing}

\safemath{\colA}{\mathscr{A}}
\safemath{\colB}{\mathscr{B}}
\safemath{\colC}{\mathscr{C}}
\safemath{\colD}{\mathscr{D}}
\safemath{\colE}{\mathscr{E}}
\safemath{\colF}{\mathscr{F}}
\safemath{\colG}{\mathscr{G}}
\safemath{\colH}{\mathscr{H}}
\safemath{\colI}{\mathscr{I}}
\safemath{\colJ}{\mathscr{J}}
\safemath{\colK}{\mathscr{K}}
\safemath{\colL}{\mathscr{L}}
\safemath{\colM}{\mathscr{M}}
\safemath{\colN}{\mathscr{N}}
\safemath{\colO}{\mathscr{O}}
\safemath{\colP}{\mathscr{P}}
\safemath{\colQ}{\mathscr{Q}}
\safemath{\colR}{\mathscr{R}}
\safemath{\colS}{\mathscr{S}}
\safemath{\colT}{\mathscr{T}}
\safemath{\colU}{\mathscr{U}}
\safemath{\colV}{\mathscr{V}}
\safemath{\colW}{\mathscr{W}}
\safemath{\colX}{\mathscr{X}}
\safemath{\colY}{\mathscr{Y}}
\safemath{\colZ}{\mathscr{Z}}

\safemath{\opA}{\mathbb{A}}
\safemath{\opB}{\mathbb{B}}
\safemath{\opC}{\mathbb{C}}
\safemath{\opD}{\mathbb{D}}
\safemath{\opE}{\mathbb{E}}
\safemath{\opF}{\mathbb{F}}
\safemath{\opG}{\mathbb{G}}
\safemath{\opH}{\mathbb{H}}
\safemath{\opI}{\mathbb{I}}
\safemath{\opJ}{\mathbb{J}}
\safemath{\opK}{\mathbb{K}}
\safemath{\opL}{\mathbb{L}}
\safemath{\opM}{\mathbb{M}}
\safemath{\opN}{\mathbb{N}}
\safemath{\opO}{\mathbb{O}}
\safemath{\opP}{\mathbb{P}}
\safemath{\opQ}{\mathbb{Q}}
\safemath{\opR}{\mathbb{R}}
\safemath{\opS}{\mathbb{S}}
\safemath{\opT}{\mathbb{T}}
\safemath{\opU}{\mathbb{U}}
\safemath{\opV}{\mathbb{V}}
\safemath{\opW}{\mathbb{W}}
\safemath{\opX}{\mathbb{X}}
\safemath{\opY}{\mathbb{Y}}
\safemath{\opZ}{\mathbb{Z}}
\safemath{\opZero}{\mathbb{O}}
\safemath{\identityop}{\opI}

\safemath{\veca}{\bma}
\safemath{\vecb}{\bmb}
\safemath{\vecc}{\bmc}
\safemath{\vecd}{\bmd}
\safemath{\vece}{\bme}
\safemath{\vecf}{\bmf}
\safemath{\vecg}{\bmg}
\safemath{\vech}{\bmh}
\safemath{\veci}{\bmi}
\safemath{\vecj}{\bmj}
\safemath{\veck}{\bmk}
\safemath{\vecl}{\bml}
\safemath{\vecm}{\bmm}
\safemath{\vecn}{\bmn}
\safemath{\veco}{\bmo}
\safemath{\vecp}{\bmp}
\safemath{\vecq}{\bmq}
\safemath{\vecr}{\bmr}
\safemath{\vecs}{\bms}
\safemath{\vect}{\bmt}
\safemath{\vecu}{\bmu}
\safemath{\vecv}{\bmv}
\safemath{\vecw}{\bmw}
\safemath{\vecx}{\bmx}
\safemath{\vecy}{\bmy}
\safemath{\vecz}{\bmz}

\safemath{\veczero}{\bmzero}
\safemath{\vecone}{\bmone}
\safemath{\vecxi}{\bmxi}
\safemath{\veclambda}{\bmlambda}
\safemath{\vecmu}{\bmmu}
\safemath{\vectheta}{\bmtheta}
\safemath{\vecphi}{\bmphi}
\safemath{\vecdelta}{\bmdelta}

\safemath{\matA}{\bA}
\safemath{\matB}{\bB}
\safemath{\matC}{\bC}
\safemath{\matD}{\bD}
\safemath{\matE}{\bE}
\safemath{\matF}{\bF}
\safemath{\matG}{\bG}
\safemath{\matH}{\bH}
\safemath{\matI}{\bI}
\safemath{\matJ}{\bJ}
\safemath{\matK}{\bK}
\safemath{\matL}{\bL}
\safemath{\matM}{\bM}
\safemath{\matN}{\bN}
\safemath{\matO}{\bO}
\safemath{\matP}{\bP}
\safemath{\matQ}{\bQ}
\safemath{\matR}{\bR}
\safemath{\matS}{\bS}
\safemath{\matT}{\bT}
\safemath{\matU}{\bU}
\safemath{\matV}{\bV}
\safemath{\matW}{\bW}
\safemath{\matX}{\bX}
\safemath{\matY}{\bY}
\safemath{\matZ}{\bZ}
\safemath{\matzero}{\bmzero}

\safemath{\matDelta}{\bDelta}
\safemath{\matLambda}{\bLambda}
\safemath{\matPhi}{\bPhi}
\safemath{\matSigma}{\bSigma}
\safemath{\matOmega}{\bOmega}
\safemath{\matTheta}{\bTheta}

\safemath{\matidentity}{\matI}
\safemath{\matone}{\matO}

\safemath{\rnda}{A}
\safemath{\rndb}{B}
\safemath{\rndc}{C}
\safemath{\rndd}{D}
\safemath{\rnde}{E}
\safemath{\rndf}{F}
\safemath{\rndg}{G}
\safemath{\rndh}{H}
\safemath{\rndi}{I}
\safemath{\rndj}{J}
\safemath{\rndk}{K}
\safemath{\rndl}{L}
\safemath{\rndm}{M}
\safemath{\rndn}{N}
\safemath{\rndo}{O}
\safemath{\rndp}{P}
\safemath{\rndq}{Q}
\safemath{\rndr}{R}
\safemath{\rnds}{S}
\safemath{\rndt}{T}
\safemath{\rndu}{U}
\safemath{\rndv}{V}
\safemath{\rndw}{W}
\safemath{\rndx}{X}
\safemath{\rndy}{Y}
\safemath{\rndz}{Z}

\safemath{\rveca}{\bimA}
\safemath{\rvecb}{\bimB}
\safemath{\rvecc}{\bimC}
\safemath{\rvecd}{\bimD}
\safemath{\rvece}{\bimE}
\safemath{\rvecf}{\bimF}
\safemath{\rvecg}{\bimG}
\safemath{\rvech}{\bimH}
\safemath{\rveci}{\bimI}
\safemath{\rvecj}{\bimJ}
\safemath{\rveck}{\bimK}
\safemath{\rvecl}{\bimL}
\safemath{\rvecm}{\bimM}
\safemath{\rvecn}{\bimN}
\safemath{\rveco}{\bomO}
\safemath{\rvecp}{\bimP}
\safemath{\rvecq}{\bimQ}
\safemath{\rvecr}{\bimR}
\safemath{\rvecs}{\bimS}
\safemath{\rvect}{\bimT}
\safemath{\rvecu}{\bimU}
\safemath{\rvecv}{\bimV}
\safemath{\rvecw}{\bimW}
\safemath{\rvecx}{\bimX}
\safemath{\rvecy}{\bimY}
\safemath{\rvecz}{\bimZ}

\safemath{\rvecxi}{\bmxi}
\safemath{\rveclambda}{\bmlambda}
\safemath{\rvecmu}{\bmmu}
\safemath{\rvectheta}{\bmtheta}
\safemath{\rvecphi}{\bmphi}

\safemath{\rmatA}{\bimA}
\safemath{\rmatB}{\bimB}
\safemath{\rmatC}{\bimC}
\safemath{\rmatD}{\bimD}
\safemath{\rmatE}{\bimE}
\safemath{\rmatF}{\bimF}
\safemath{\rmatG}{\bimG}
\safemath{\rmatH}{\bimH}
\safemath{\rmatI}{\bimI}
\safemath{\rmatJ}{\bimJ}
\safemath{\rmatK}{\bimK}
\safemath{\rmatL}{\bimL}
\safemath{\rmatM}{\bimM}
\safemath{\rmatN}{\bimN}
\safemath{\rmatO}{\bimO}
\safemath{\rmatP}{\bimP}
\safemath{\rmatQ}{\bimQ}
\safemath{\rmatR}{\bimR}
\safemath{\rmatS}{\bimS}
\safemath{\rmatT}{\bimT}
\safemath{\rmatU}{\bimU}
\safemath{\rmatV}{\bimV}
\safemath{\rmatW}{\bimW}
\safemath{\rmatX}{\bimX}
\safemath{\rmatY}{\bimY}
\safemath{\rmatZ}{\bimZ}

\safemath{\rmatDelta}{\bimDelta}
\safemath{\rmatLambda}{\bimLambda}
\safemath{\rmatPhi}{\bimPhi}
\safemath{\rmatSigma}{\bimSigma}
\safemath{\rmatOmega}{\bimOmega}
\safemath{\rmatTheta}{\bimTheta}

\usepackage{amssymb}
\usepackage{amsfonts}
\usepackage{mathrsfs}
\usepackage{xspace}
\usepackage{bm}
\usepackage{fancyref}
\usepackage{textcomp}

\usepackage{multirow}
\usepackage{stmaryrd}

\newenvironment{textbmatrix}{	\setlength{\arraycolsep}{2.5pt}%
								\big[\begin{matrix}}{\end{matrix}\big]%
								\raisebox{0.08ex}{\vphantom{M}}}

\def\be{\begin{equation}}
\def\ee{\end{equation}}
\def\een{\nonumber \end{equation}}
\def\mat{\begin{bmatrix}}
\def\emat{\end{bmatrix}}
\def\btm{\begin{textbmatrix}}
\def\etm{\end{textbmatrix}}

\def\ba#1\ea{\begin{align}#1\end{align}}
\def\bas#1\eas{\begin{align*}#1\end{align*}}
\def\bs#1\es{\begin{split}#1\end{split}} 
\def\bg#1\eg{\begin{gather}#1\end{gather}}
\def\bml#1\eml{\begin{multline}#1\end{multline}}
\def\bi#1\ei{\begin{itemize}#1\end{itemize}}

\safemath{\dirac}{\delta}					
\safemath{\krond}{\dirac}					

\safemath{\upto}{\uparrow}
\safemath{\downto}{\downarrow}
\safemath{\iu}{j}							
\safemath{\ev}{\lambda}						
\safemath{\hilseqspace}{l^{2}}				
\newcommand{\banachfunspace}[1]{\setL^{#1}}	
\safemath{\hilfunspace}{\banachfunspace{2}}	

\safemath{\SNR}{\textsf{SNR}} 				
\safemath{\PAR}{\textsf{PAR}} 				
\safemath{\No}{N_0}							
\safemath{\Es}{E_s}							
\safemath{\Eb}{E_b}							
\safemath{\EbNo}{\frac{\Eb}{\No}}
\safemath{\EsNo}{\frac{\Es}{\No}}

\DeclareMathOperator{\CHop}{\ensuremath{\opH}} 
\safemath{\tvir}{\rndh_{\CHop}}				
\safemath{\tvtf}{\rndl_{\CHop}}				
\safemath{\spf}{\rnds_{\CHop}}				
\safemath{\bff}{H_{\CHop}}					

\safemath{\ircf}{r_{h}}						
\safemath{\tftvcf}{r_{s}}					
\safemath{\tfcf}{r_{l}}						
\safemath{\bfcf}{r_{H}}						

\safemath{\tcorr}{c_h}						
\safemath{\scf}{c_{s}}						
\safemath{\tfcorr}{c_{l}}					
\safemath{\fcorr}{c_{H}}						

\safemath{\mi}{I}							
\safemath{\capacity}{C}						

\safemath{\normal}{\mathcal{N}}			
\safemath{\jpg}{\mathcal{CN}}			
\safemath{\mchain}{\leftrightarrow}		

\safemath{\dB}{\,\mathrm{dB}}
\safemath{\dBm}{\,\mathrm{dBm}}
\safemath{\Hz}{\,\mathrm{Hz}}
\safemath{\kHz}{\,\mathrm{kHz}}
\safemath{\MHz}{\,\mathrm{MHz}}
\safemath{\GHz}{\,\mathrm{GHz}}
\safemath{\s}{\,\mathrm{s}}
\safemath{\ms}{\,\mathrm{ms}}
\safemath{\mus}{\,\mathrm{\text{\textmu}s}}
\safemath{\ns}{\,\mathrm{ns}}
\safemath{\ps}{\,\mathrm{ps}}
\safemath{\meter}{\,\mathrm{m}}
\safemath{\mm}{\,\mathrm{mm}}
\safemath{\cm}{\,\mathrm{cm}}
\safemath{\m}{\,\mathrm{m}}
\safemath{\W}{\,\mathrm{W}}
\safemath{\mW}{\, \mathrm{mW}}
\safemath{\J}{\,\mathrm{J}}
\safemath{\K}{\,\mathrm{K}}
\safemath{\bit}{\,\mathrm{bit}}
\safemath{\nat}{\,\mathrm{nat}}

\safemath{\define}{\triangleq}			

\safemath{\equivalent}{\sim}
\safemath{\distas}{\sim}					
\safemath{\sdiff}{\Delta}				

\safemath{\reals}{\mathbb{R}}
\safemath{\positivereals}{\reals_{+}}
\safemath{\integers}{\mathbb{Z}}
\safemath{\posint}{\integers_{+}}
\safemath{\naturals}{\mathbb{N}}
\safemath{\posnaturals}{\naturals_{+}}
\safemath{\complexset}{\mathbb{C}}
\safemath{\rationals}{\mathbb{Q}}

\newcommand*{\fancyrefapplabelprefix}{app}		
\newcommand*{\fancyrefthmlabelprefix}{thm}		
\newcommand*{\fancyreflemlabelprefix}{lem}		
\newcommand*{\fancyrefcorlabelprefix}{cor}		
\newcommand*{\fancyrefdeflabelprefix}{def}		
\newcommand*{\fancyrefproplabelprefix}{prop}	
\newcommand*{\fancyrefobslabelprefix}{obs}		
\newcommand*{\fancyrefalglabelprefix}{alg}		
\newcommand*{\fancyrefasmlabelprefix}{asm}	    
\newcommand*{\fancyreftbllabelprefix}{tbl}	    

\frefformat{vario}{\fancyrefseclabelprefix}{Section~#1}
\frefformat{vario}{\fancyrefthmlabelprefix}{Theorem~#1}
\frefformat{vario}{\fancyreflemlabelprefix}{Lemma~#1}
\frefformat{vario}{\fancyrefcorlabelprefix}{Corollary~#1}
\frefformat{vario}{\fancyrefdeflabelprefix}{Definition~#1}
\frefformat{vario}{\fancyrefobslabelprefix}{Observation~#1}
\frefformat{vario}{\fancyrefasmlabelprefix}{Assumption~#1}
\frefformat{vario}{\fancyreffiglabelprefix}{Figure~#1}
\frefformat{vario}{\fancyrefapplabelprefix}{Appendix~#1} 
\frefformat{vario}{\fancyrefproplabelprefix}{Proposition~#1}
\frefformat{vario}{\fancyrefalglabelprefix}{Algorithm~#1}
\frefformat{vario}{\fancyrefeqlabelprefix}{(#1)}
\frefformat{vario}{\fancyreftbllabelprefix}{Table~#1}

\safemath{\dictab}{[\,\dicta\,\,\dictb\,]}

\safemath{\ysig}{\bmy}
\safemath{\ysighat}{\hat{\ysig}}
\safemath{\ysigdim}{M}
\safemath{\xsig}{\bmx}
\safemath{\xsigdim}{N}
\safemath{\nx}{n_x}
\safemath{\zsig}{\bmz}
\safemath{\zsigdim}{\ysigdim}
\safemath{\rsig}{\bmr}
\safemath{\Adict}{\bA}
\safemath{\Adicttilde}{\widetilde{\Adict}}
\safemath{\Adictdim}{\outputdim\times\xsigdim}
\safemath{\avec}{\bma}
\safemath{\avectilde}{\tilde{\avec}}
\safemath{\Bdict}{\bB}
\safemath{\Bdicttilde}{\widetilde{\Bdict}}
\safemath{\Cdict}{\bC}
\safemath{\cvec}{\bmc}
\safemath{\Ddict}{\bD}
\safemath{\Ddictdim}{\ysigdim\times\xsigdim}
\safemath{\dvec}{\bmd}
\safemath{\Ddicttilde}{\widetilde{\bD}}
\safemath{\Bonb}{\bB}
\safemath{\bvec}{\bmb}
\safemath{\Bonbdim}{\ysigdim\times\ysigdim}
\safemath{\noise}{\bmn}
\safemath{\noisedim}{\ysigim}
\safemath{\err}{\bme}
\safemath{\errdim}{\ysigdim}
\safemath{\errset}{\setE}
\safemath{\nerr}{n_e}
\safemath{\delop}{\bP_\errset}
\safemath{\delopc}{\bP_{{\errset}^c}}

\safemath{\cplxi}{\imath}
\safemath{\cplxj}{\jmath}

\safemath{\dict}{\matD}
\safemath{\inputdim}{N}		
\safemath{\outputdim}{M}		
\safemath{\sparsity}{S}	
\safemath{\inputdimA}{{N_a}}	
\safemath{\inputdimB}{{N_b}}	
\safemath{\elemA}{{n_a}}	
\safemath{\elemB}{{n_b}}	
\safemath{\resA}{\matR_a}	
\safemath{\resB}{\matR_b}	
\safemath{\subD}{\matS} 
\safemath{\subA}{\matS_a} 
\safemath{\subB}{\matS_b} 
\safemath{\dicta}{\matA} 	
\safemath{\dictb}{\matB} 	
\safemath{\hollowS}{H}
\safemath{\hollowA}{H_a}
\safemath{\hollowB}{H_b}
\safemath{\cross}{Z}
\safemath{\coh}{\mu_d}			
\safemath{\coha}{\mu_a}			
\safemath{\cohb}{\mu_b}			
\safemath{\mubs}{\nu}	
\safemath{\cohm}{\mu_m} 
\safemath{\dictset}{\setD}	
\safemath{\dictsetp}{\dictset(\coh,\coha,\cohb)}	
\safemath{\dictsetgen}{\dictset_\text{gen}}
\safemath{\dictsetgenp}{\dictsetgen(\coh)}
\safemath{\dictsetonb}{\dictset_\text{onb}}
\safemath{\dictsetonbp}{\dictsetonb(\coh)}

\safemath{\leftside}{U}
\safemath{\rightsideA}{R_a}
\safemath{\rightsideB}{R_b}

\safemath{\indexS}{\setI_S} 

\safemath{\na}{n_a}			
\safemath{\nb}{n_b}			
\safemath{\coeffa}{p_i}	
\safemath{\coeffb}{q_j}	
\safemath{\seta}{\setP}		
\safemath{\setb}{\setQ}     
\safemath{\setw}{\setW}	
\safemath{\setz}{\setZ}	
\safemath{\cola}{\veca}		
\safemath{\colb}{\vecb}		
\safemath{\cold}{\vecd}		
\safemath{\inputvec}{\vecx} 	
\safemath{\error}{\vece}	
\safemath{\noiseout}{\vecz} 	
\safemath{\inputvecel}{x}
\safemath{\inputveca}{\vecx_a}
\safemath{\inputvecb}{\vecx_b}
\safemath{\outputvec}{\vecy}	
\safemath{\lambdamin}{\lambda_{\mathrm{min}}}

\safemath{\elltwo}{\ell_2}
\safemath{\ellone}{\ell_1}
\safemath{\ellzero}{\ell_0}
\safemath{\ellinf}{\ell_\infty}
\safemath{\ellinftilde}{\ell_{\widetilde\infty}}
\safemath{\licard}{Z(\coh,\coha,\cohb)}
\safemath{\xsol}{\hat{x}}
\safemath{\xbord}{x_b}		
\safemath{\xstat}{x_s}		
\safemath{\xstatLone}{\tilde{x}_s}
\safemath{\order}{\mathcal{O}} 
\safemath{\scales}{\Theta} 
\safemath{\ones}{\mathbf{1}} 
\safemath{\zeroes}{\mathbf{0}} 
\safemath{\thlone}{\kappa(\coh,\cohb)} 
\safemath{\constoneA}{\delta} 
\safemath{\constoneB}{\epsilon} 
\safemath{\nlarge}{L}				   
\safemath{\sumlarge}{S_\nlarge}
\safemath{\maxlarger}{P_\nlarge}	   
\safemath{\Pzero}{\textrm{P0}}	
\safemath{\Pone}{\textrm{P1}}
\safemath{\vecfir}{\vecw}			 
\safemath{\vecsec}{\vecz}
\safemath{\elvecfir}{w}              
\safemath{\elvecsec}{z}				 
\safemath{\nlargefir}{n}
\safemath{\normout}{\gamma}
\safemath{\auxfun}{h}
\safemath{\supp}{\textrm{supp}}

\safemath{\indexa}{\ell}
\safemath{\indexb}{r}
\safemath{\indexc}{i}
\safemath{\indexd}{j}

\safemath{\project}{P}

\IEEEoverridecommandlockouts

\safemath{\Herm}{\textnormal{H}}

\begin{document}

\bstctlcite{IEEEexample:BSTcontrol}

\title{
A Resolution-Adaptive  8\,\text{mm}$^\text{2}$ 9.98\,Gb/s 39.7\,pJ/b 32-Antenna All-Digital Spatial Equalizer for mmWave Massive MU-MIMO in 65nm CMOS}
\author{\IEEEauthorblockN{Oscar Casta\~neda$^1$,  Zachariah Boynton$^2$, Seyed Hadi Mirfarshbafan$^1$, \\ Shimin Huang$^2$, Jamie C. Ye$^2$, Alyosha Molnar$^2$, and Christoph Studer$^1$} \\[-0.3cm]
\thanks{This work was supported in part by  ComSenTer, one of six centers in JUMP, a SRC program sponsored by DARPA. The work of CS was also supported by an ETH Research Grant and by the US NSF under grants CNS-1717559 and ECCS-1824379. Contact author: O. Casta\~neda (e-mail: caoscar@ethz.ch)}
\IEEEauthorblockA{\small $^1$Department of Information Technology and Electrical Engineering, ETH Z\"urich, Switzerland \\ $^2$School of Electrical and Computer Engineering, Cornell University, Ithaca, NY} 
}

\maketitle

\begin{abstract}
All-digital millimeter-wave (mmWave) massive multi-user multiple-input multiple-output (MU-MIMO) receivers enable extreme data rates but require high power consumption. 
In order to reduce power consumption, this paper presents the first resolution-adaptive all-digital receiver ASIC that is able to adjust the resolution of the data-converters and baseband-processing engine to the instantaneous communication scenario.
The scalable 32-antenna, 65\,nm CMOS receiver occupies a total area of 8\,mm$^\text{2}$ and integrates analog-to-digital converters (ADCs) with programmable gain and resolution, beamspace channel estimation, and a resolution-adaptive processing-in-memory spatial equalizer.
With 6-bit ADC samples and a 4-bit spatial equalizer, our ASIC achieves a throughput of 9.98\,Gb/s while being at least $\bf2\boldsymbol\times$ more energy-efficient than state-of-the-art designs.
\end{abstract}


\section{Introduction}\label{sec:intro}

Future wireless communication systems are expected to operate at millimeter-wave (mmWave) frequencies to exploit large amounts of unoccupied contiguous bandwidth~\cite{rappaport15a}.
Multiple-input multiple-output (MIMO) can be utilized to combat the strong path loss at mmWave frequencies while enabling  improvements in spectral efficiency by supporting multi-user (MU) communication~\cite{larsson14a}.
However, the concoction of wide bandwidths and large antenna arrays at the basestation (BS) results in a number of serious implementation challenges.

Power consumption is a major concern in the design of such mmWave massive MU-MIMO systems.
While hybrid analog-digital BS architectures have been proposed to address this issue~\cite{alkhateeb14a}, they are limited in their spatial-multiplexing capabilities resulting in a spectral efficiency loss.
In contrast, all-digital BS architectures, with one radio-frequency (RF) chain and dedicated analog-to-digital converters (ADCs) for each antenna, are able to realize the full potential of massive MU-MIMO while achieving comparable RF power consumption to hybrid architectures~\cite{roth2017achievable,panagiotis20}.
However, in order to achieve comparable power consumption to hybrid solutions, all-digital BS architectures must rely on low-resolution ADCs~\cite{jacobsson17b} and low-resolution digital baseband processing~\cite{castaneda19fame}.
As it has been shown in~\cite{yan2019performance,abdelghany18}, the required resolution of ADCs and baseband processing depends on the system conditions, such as the number of user equipments (UEs), the modulation scheme, and the channel propagation scenario.
This implies that energy-efficient all-digital BS architectures should dynamically adapt the ADC and baseband-processing resolutions to the instantaneous communication scenario, as opposed to hard-wiring those parameters for the worst case during design time. 

\subsubsection*{Contributions}
We propose the first \textit{resolution-adaptive} ASIC for all-digital mmWave massive MU-MIMO BSs. 
In order to maximize energy efficiency, our design is able to dynamically adapt the ADC and baseband-processing resolutions to the instantaneous communication scenario.
The 8\,mm$^\text{2}$ 65\,nm CMOS ASIC supports 32 BS antennas and up to 16 UEs. 
The ASIC tightly integrates an array of time-interleaved (TI) successive-approximation register (SAR) ADCs, a beamspace channel estimation (BEACHES) engine, and a spatial equalizer. 
To operate at low resolution while achieving high spectral efficiency, the spatial equalizer builds upon finite-alphabet equalization~\cite{castaneda19fame}, an emerging baseband-processing paradigm that is implemented using an energy-efficient standard-cell-based processing-in-memory (PIM) architecture.
We provide measurement results to demonstrate that 
our resolution-adaptive ASIC outperforms existing equalizers in terms of energy and area efficiency while achieving highest-in-class throughput.


\section{mmWave Massive MU-MIMO} \label{sec:system}

\subsection{System Model}

\begin{figure}[tp]
\centering
\includegraphics[width=0.98\columnwidth]{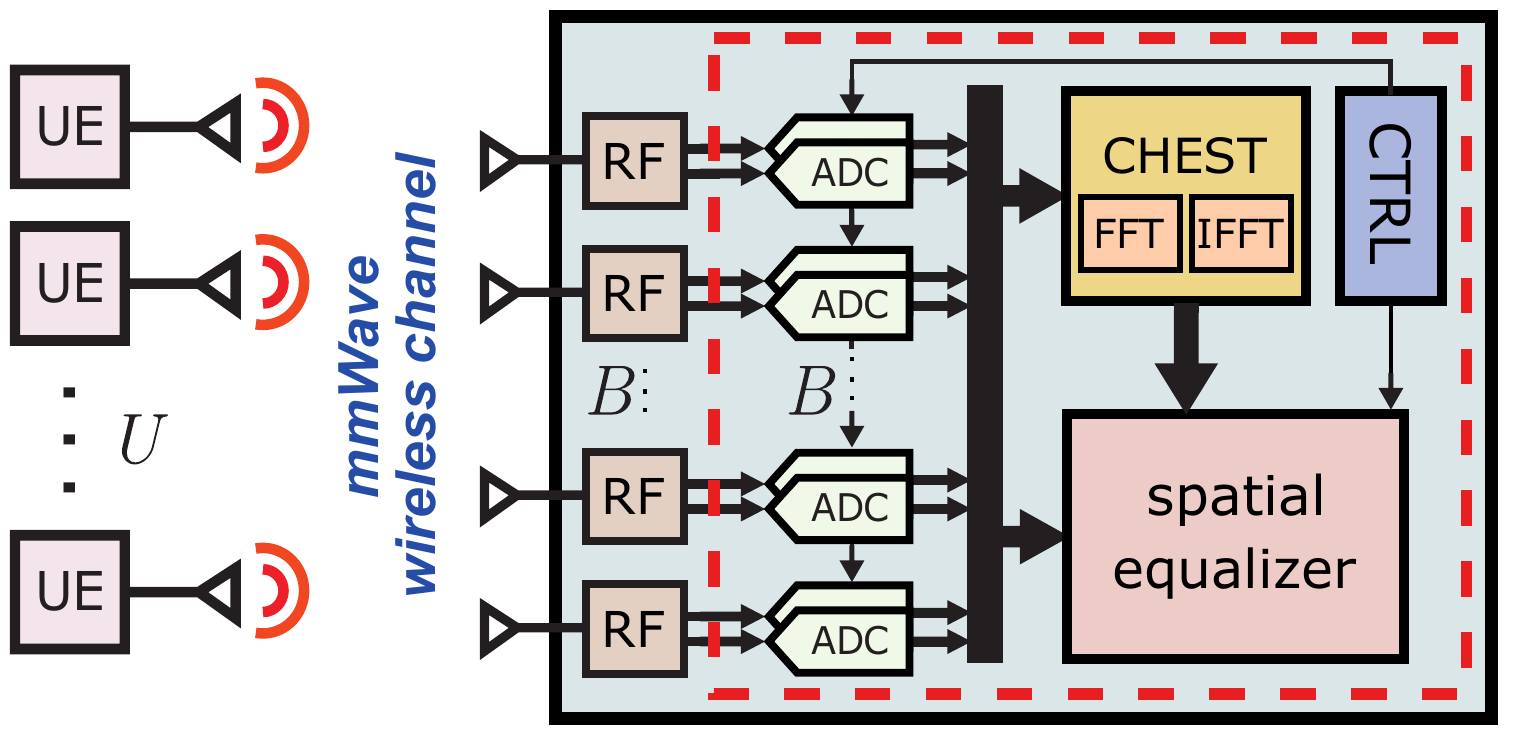}
\vspace{-0.15cm}
\caption{Overview of the considered all-digital mmWave massive MU-MIMO system in which $U$ UEs communicate with a $B$-antenna BS. Our ASIC includes the parts within the dashed red rectangle: an ADC array, a channel estimation (CHEST) engine, and a spatial equalizer. The resolutions of the ADCs and equalizer can be adapted to the communication scenario.}
\label{fig:system_overview}
\end{figure}

We consider the mmWave massive MU-MIMO uplink, as illustrated in \fref{fig:system_overview}, where $U$ single-antenna UEs transmit data to a $B$-antenna BS.
The baseband channel is modeled using the frequency-flat input-output relation $\bmy=\bH\bms+\bmn$, where $\bmy\in\complexset^B$ contains the signals received at the BS antennas, $\bH\in\complexset^{B\times U}$ is the MIMO channel matrix, $\bms\in\setS^U$ contains the UE-transmit symbols taken from the constellation $\setS$ (e.g., $16$-QAM), and $\bmn\in\complexset^B$ is complex Gaussian noise with variance $\No$ per entry.
The UE-transmit symbols $s_u$, $u=1,\dots,U$, are zero mean with variance $\Es\sigma^2_u$, and we assume $\pm3\,$dB UE-side power control so that $\max_{u}\{\sigma^2_u\|\bmh_u\|_2^2\}/\min_{u}\{\sigma^2_u\|\bmh_u\|_2^2\}=4$ with $\bmh_u$ being the $u$th column of $\bH$~\cite{castaneda19fame}.
The received vector $\bmy$ is quantized by the ADCs resulting in $\bmz$, which is then used to estimate $\bH$ and generate estimates $\hat{\bms}$ of the transmit vector $\bms$.

\subsection{Beamspace Channel Estimation}
The channel estimate $\widetilde{\bH}$ is generated during a training phase.
Each UE transmits a pilot symbol $\phi\in\setS$ known to the BS, while the other UEs remain silent, resulting in $\bmy = \phi\bmh + \bmn$, where $\bmh$ is the channel vector of the pilot-transmitting UE.
The least squares (LS) estimate of the channel vector is then computed as $\tilde{\bmh} = \bmz/\phi$.
The LS estimate can then be transformed into the so-called \emph{beamspace domain} by taking the discrete Fourier transform (DFT) of $\tilde{\bmh}$.
Since beamspace channel vectors are typically sparse in mmWave systems~\cite{rappaport15a}, we can use the BEACHES (short for beamspace channel estimation) algorithm~\cite{mirfarshbafan19a} to denoise the channel vectors, resulting in  significant performance improvements.

\subsection{Finite-Alphabet Equalization}
At the extremely high bandwidths offered by mmWave frequencies, even linear spatial equalizers of the form $\hat{\bms}=\bW^\Herm\bmz$ result in power-hungry hardware~\cite{castaneda19fame}.
To minimize power consumption without sacrificing  spectral efficiency, finite-alphabet equalization~\cite{castaneda19fame} proposes to allow per-UE scaling factors (contained in a vector $\boldsymbol\mu$), so that the entries of the equalization matrix $\bW^\Herm$ can be represented with low-resolution numbers in $\bX^\Herm$  (e.g., 5 bits or lower).
With an equalization matrix of the form  $\bV^\Herm\!=\!\mathrm{diag}\left(\boldsymbol\mu\right)\bX^\Herm$, the main complexity of equalization now lies in computing $\bX^\Herm\bmz$, which can be implemented with low power digital circuitry~\cite{castaneda19fame,castaneda20tcas}.


\section{VLSI architecture} \label{sec:arch}

\fref{fig:system_overview} shows the top-level architecture of our $B\!=\!32$ antenna all-digital receiver ASIC. 
The design contains an analog front-end with $2B\!=\!64$ ADCs to quantize the I/Q receive vector~$\bmy$ into $\bmz$.
The vector $\bmz$ is used to estimate the symbols $s_u$ transmitted by the $U\!\leq\!16$ UEs with a specialized architecture of the parallel processor in associative content-addressable-memory (PPAC) \cite{castaneda2019ppac}.
The ADCs are programmable to support different resolutions in $\bmz$, while PPAC supports different resolutions in $\bX^\Herm$ and in $\bmz$.
The quantized vector $\bmz$ can also be used to extract channel estimates using the BEACHES engine.

As detailed in \fref{sec:arch_eq}, the PPAC spatial equalizer operates bit-serially, i.e., it only processes one bit from all $\bmz$-entries per clock cycle.
Since PPAC only operates one bit per $\bmz$-entry at a time, it can be tightly coupled with a SAR ADC that converts one bit per $\bmz$-entry at a time.
As a single instance of PPAC is not sufficient to support the high bandwidths offered by mmWave systems, especially when implemented in a 65\,nm technology node, we deployed four PPAC instances.
The ADCs are TI by the same factor of four so that each PPAC instance is tightly coupled with its own ADC array.
This co-design of ADC array and spatial equalizer enables one to easily scale our system to larger bandwidths with (i) more silicon area and/or (ii) a more advanced technology node, which would also significantly increase the per-instance throughput.
Moreover, both the ADC array and PPAC can be scaled to more BS antennas or UEs.
However, our current ASIC is limited to $B=32$ and $U\leq16$ due to silicon area and pin-I/O constraints.

\subsection{ADC Array} \label{sec:arch_afe}

\begin{figure}
\centering
\includegraphics[width=0.98\columnwidth]{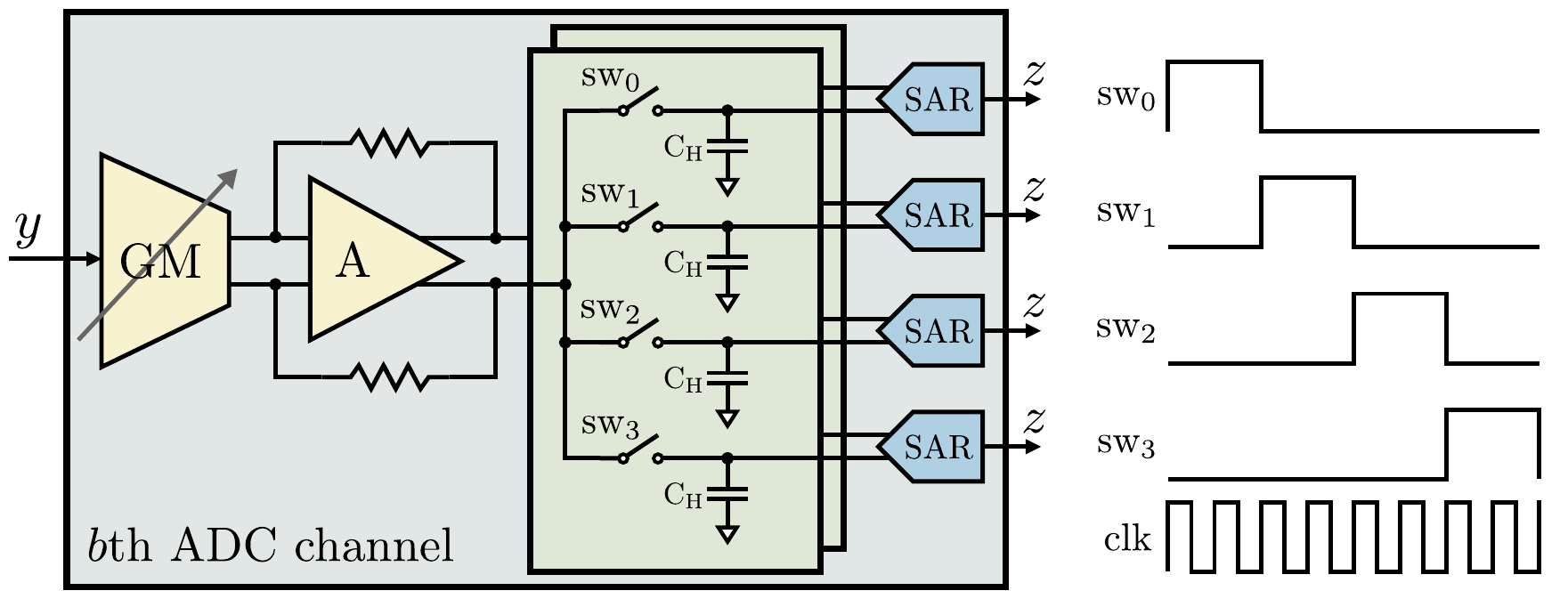}
\vspace{-0.15cm}
\caption{Block diagram of an ADC channel and clocks for 6-bit samples.}
\label{fig:ADC_block}
\end{figure}

The ADC array comprises $2B\!=\!64$ channels (32 I/Q channels).
As shown in \fref{fig:ADC_block}, each ADC channel consists of a two-stage programmable gain amplifier (PGA) and four TI-SAR ADCs.
The two-stage PGA provides programmable gain, which can be set to amplifications from 1$\times$ to 32$\times$ in powers of two.
The first stage of the PGA is a cascode amplifier with switchable transconductance cells to provide programmable gain; the second stage is a fixed-gain inverter-based amplifier. 
Sampling switches, controlled by non-overlapping sampling clocks, are used to TI the four fully-differential SAR ADCs.

The ADC array quantizes $\bmy$ into $\bmz$ using mid-rise quantization with either 3 bits or 6 bits of resolution:
For the \{3,6\}-bit case, \{1,2\} clock cycles are used for sampling, requiring a total of \{4,8\} clock cycles per sample.
During the sampling clock cycles, the associated PPAC instance is idle.
Reducing the ADC resolution either enables higher sampling rates (and, hence, higher throughput) or lowers power consumption.

\subsection{PPAC-Based Spatial Equalizer} \label{sec:arch_eq}

\begin{figure}[tp]
\centering
\includegraphics[width=0.9\columnwidth]{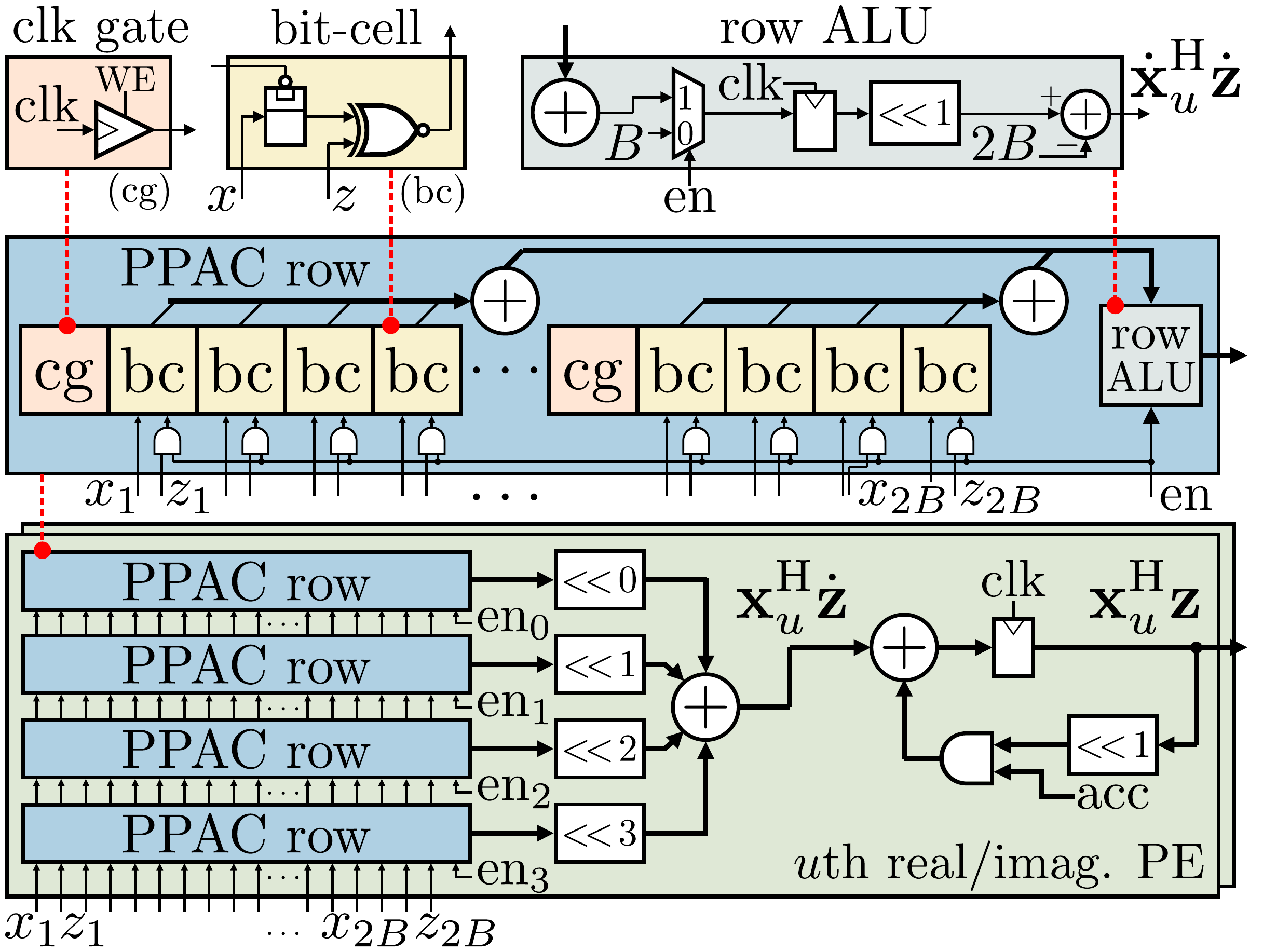}
\vspace{-0.15cm}
\caption{Block diagram of the specialized PPAC for mmWave spatial equalization with mid-rise inputs. The architecture uses one PPAC row to compute an inner-product between two 1-bit vectors, $\dot{\bmx}_u$ and $\dot{\bmz}$. Multi-bit vectors $\bmx_u$ and $\bmz$ are processed using spatial replication and bit-serial operation, respectively.}
\label{fig:arch_eq}
\end{figure}

\fref{fig:arch_eq} illustrates our spatial equalizer, which is a  version of the PPAC architecture~\cite{castaneda19fame} specialized for spatial equalization.
PPAC is an all-digital, standard-cell-based PIM architecture for matrix-vector-product-like tasks, in which each memory bit-cell contains both storage (a latch) and logic (an XNOR).
PPAC uses a so-called PPAC row to compute an inner-product between two vectors with one-bit entries.
To support multi-bit operation, spatial replication is used for one of the vectors (a row of $\bX^\Herm)$ and bit-serial operation for the other ($\bmz$).
Our PPAC implementation supports 1 bit to 4 bits of resolution in $\bX^\Herm$; the inputs to each PPAC row can be muted to save power when operating with fewer than 4 bits per $\bX^\Herm$-entry.
It also supports up to 8 bits of resolution in $\bmz$: To operate a $q$-bit $\bmz$, a PPAC instance needs to run for $q$ clock cycles.

The results in~\cite{castaneda20tcas} have shown that the PIM-based PPAC can achieve lower power consumption than a conventional, non-PIM architecture when performing equalization with low-resolution matrices.
In contrast to~\cite{castaneda20tcas}, our PPAC implementation operates with mid-rise quantizers, is programmable in the number of equalization matrix $\bX^\Herm$ bits, and is tightly coupled to the ADC array.
In addition, this is the first silicon realization of a PPAC-based hardware accelerator. 

\begin{figure}[tp]
\centering
\includegraphics[width=0.98\columnwidth]{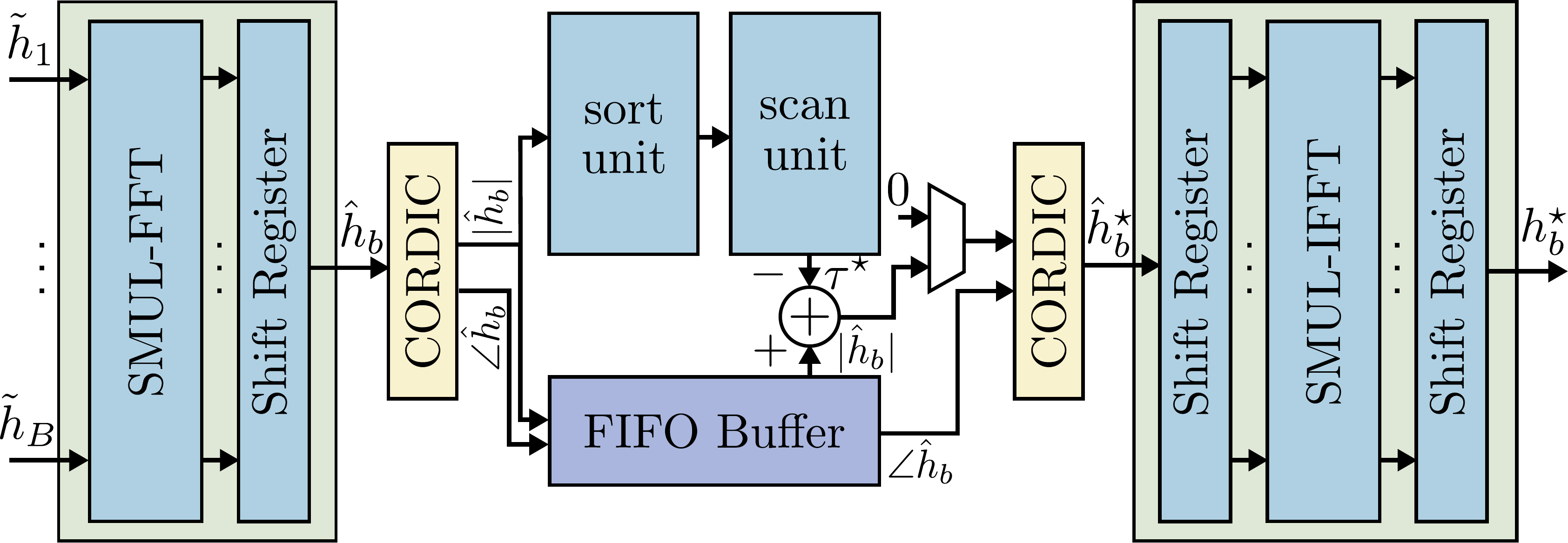}
\vspace{-0.15cm}
\caption{Block diagram of the BEACHES engine. A streaming multiplierless (SMUL) FFT and IFFT are used for the beamspace transform. The sort and scan units determine the MSE-optimal denoising threshold $\tau^\star$, which is applied on the magnitudes (extracted with a CORDIC) of the beamspace channel vector.}
\label{fig:arch_bx}
\end{figure}

\subsection{BEACHES Engine}
BEACHES~\cite{mirfarshbafan19a} is a mmWave channel denoising algorithm that performs soft-thresholding on the magnitudes of the estimated beamspace channel vector.
The algorithm relies on Stein's unbiased risk estimator to apply a near mean-square error (MSE)-optimal denoising threshold.
BEACHES uses a fast Fourier transform (FFT) to transform the channel estimate $\tilde{\bmh}$ to  beamspace $\hat{\bmh}$.
The denoised vector $\hat{\bmh}^\star$ is then transformed back to the antenna-domain with an inverse FFT~(IFFT).
Our BEACHES implementation is adopted from the architecture in~\cite{mirfarshbafan19a} and illustrated in \fref{fig:arch_bx}.
Unlike~\cite{mirfarshbafan19a}, we use a streaming multiplierless (SMUL) FFT architecture from~\cite{mirfarshbafan21} for both the FFT and IFFT.
Our design is capable of denoising a new channel vector $\tilde{\bmh}$ every $B=32$ clock cycles.


\section{ASIC Implementation Results}\label{sec:impl}

\begin{figure}[tp]
\centering
\subfigure[BER vs.\ SNR]{\includegraphics[width=.48\columnwidth]{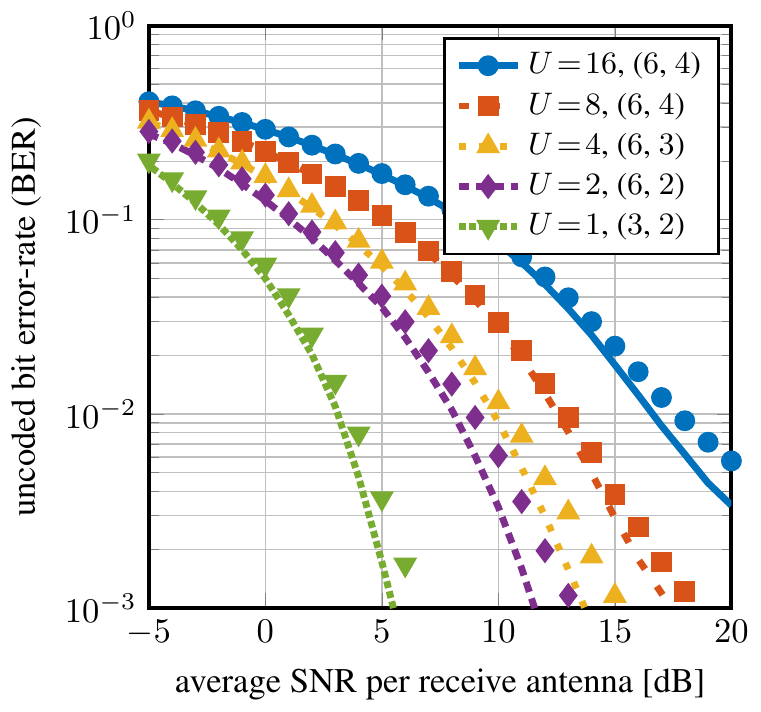}\label{fig:perf_ber}}
\hfill
\subfigure[Energy vs.\ SNR loss  ]{\includegraphics[width=.46\columnwidth]{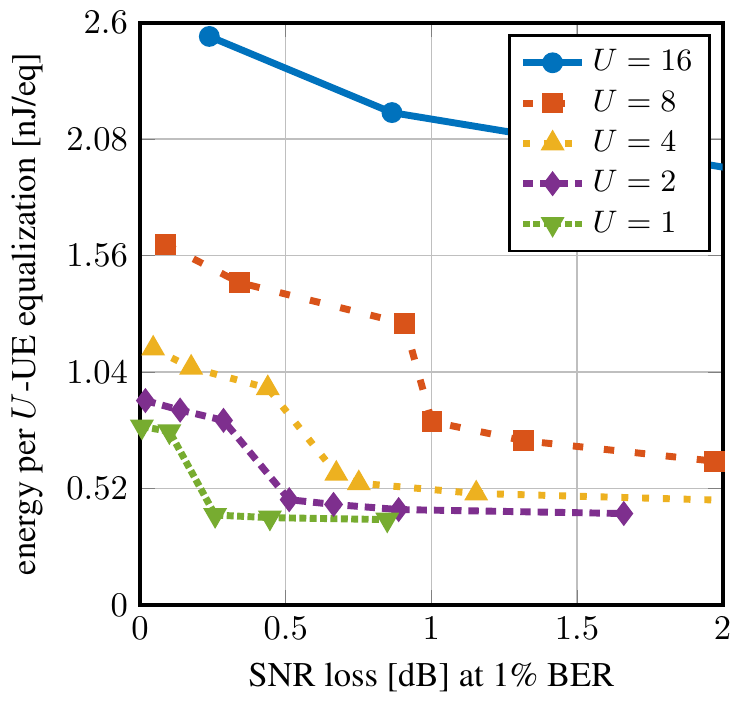}\label{fig:perf_energy_ber}}
\caption{Performance of the implemented 32-antenna ASIC over a LoS mmMAGIC UMi mmWave channel in terms of (a) bit-error rate (BER) with 16-QAM and (b) ADC array and spatial equalizer energy with QPSK. In (a), lines correspond to the performance of infinite-resolution ADCs and L-MMSE equalizer, markers to the ADC and equalizer resolution given in parentheses.}\label{fig:perf}
\end{figure}

\begin{figure}[tp]
\centering
\includegraphics[height=4.2cm]{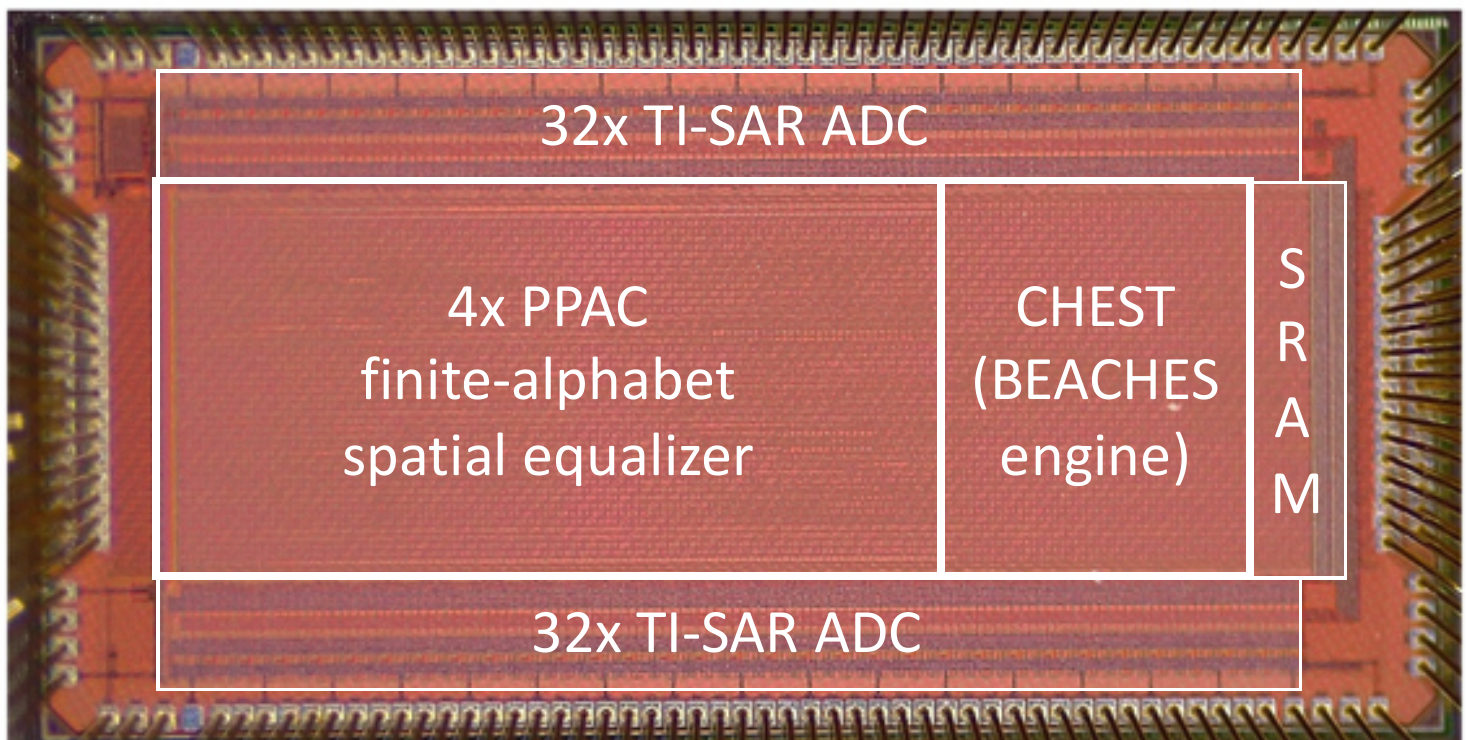}
\caption[caption]{Micrograph of the 8\,mm$^2$ all-digital receiver ASIC in 65\,nm CMOS.}
\label{fig:chip}
\end{figure}

\begin{figure}
\centering
\subfigure[PPAC spatial equalizer]{\includegraphics[width=0.48\columnwidth]{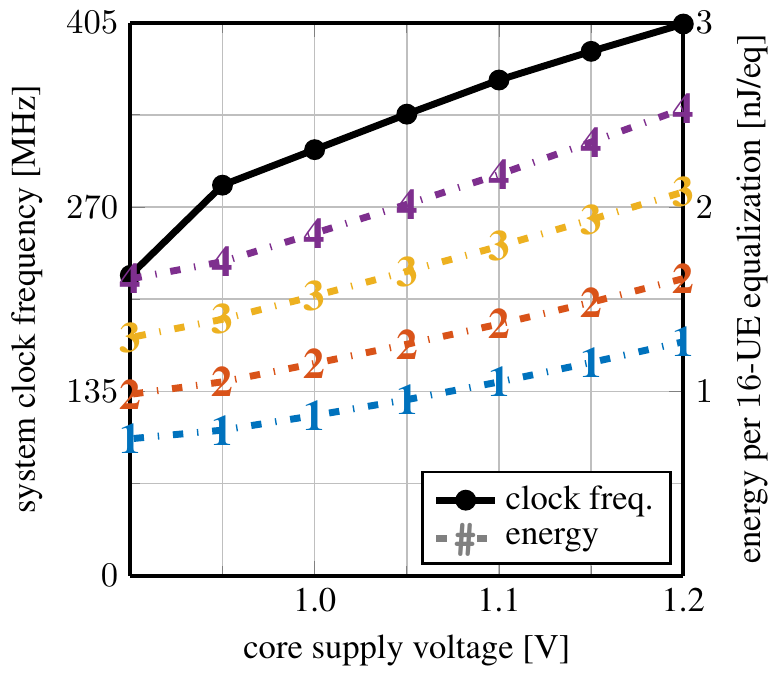}\label{fig:vfs_ppac}}
\hfill
\subfigure[BEACHES engine]{\includegraphics[width=0.48\columnwidth]{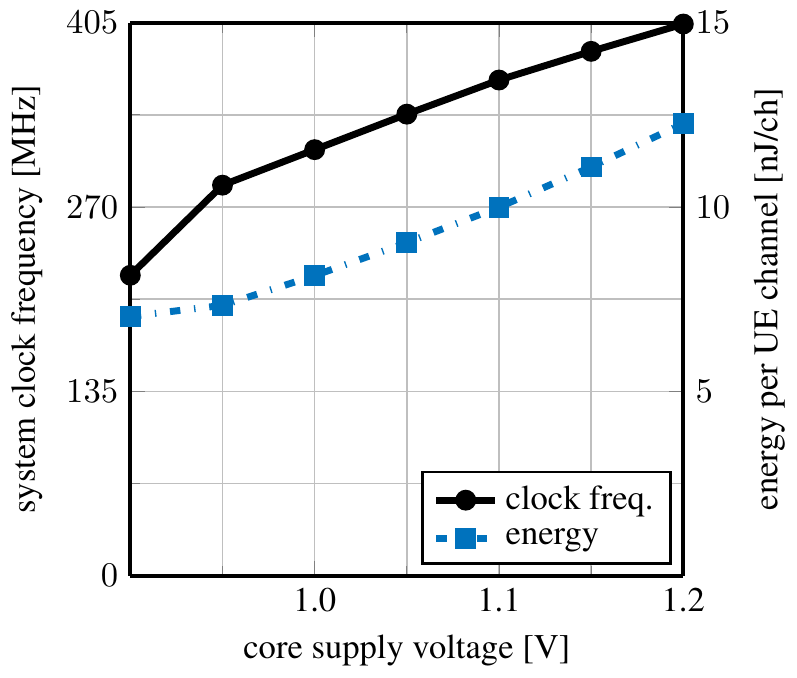}\label{fig:vfs_beaches}}
\caption{System clock frequency and energy versus the supply voltage for the (a) PPAC spatial equalizer and (b)  BEACHES engine. The ADC resolution is kept constant at $6$ bits and the PPAC resolution is indicated by the markers.}
\label{fig:vfs}
\end{figure}

\subsection{Bit Error-Rate (BER) Performance}

\fref{fig:perf_ber} shows the uncoded BER of our $B=32$ receiver ASIC for different UE numbers~$U$ with line-of-sight (LoS) mmMAGIC UMi channels generated at a carrier frequency of 60\,GHz using QuaDRiGA~\cite{jaeckel2014quadriga}.
Our ASIC's performance (markers) achieves that of an infinite-resolution linear minimum MSE (L-MMSE) equalizer (lines) within $1$\,dB SNR loss measured at $1\%$ BER.
For common massive MU-MIMO load factors $\beta= U/B \leq 1/8$ ($U\leq4$), where linear equalization achieves near-optimal BER performance, our ASIC can operate at reduced ADC and equalization resolutions.

The ADC and equalizer resolutions affect both the ASIC's error-rate performance and energy efficiency, as illustrated in \fref{fig:perf_energy_ber} for QPSK transmission and LoS channels:
Markers represent the per-equalization energy and performance (in terms of SNR loss at $1\%$ BER) for Pareto-optimal configurations of ADC and equalizer resolution; e.g., the left-most marker of each curve corresponds to the maximum-resolution configuration of $6$ ADC bits and $4$ equalizer bits. 
\fref{fig:perf_energy_ber} shows that for an SNR loss of $0.5\,$dB, lower load factors require lower resolution, which yields lower per-equalization energy, saving up to $2\times$ energy compared to the maximum-resolution configuration.
In situations that allow for higher SNR losses, the resolutions and, hence, the energy, can be reduced even further. 

\subsection{ASIC Measurements and Comparison}
\fref{fig:chip} shows the 8\,mm$^2$ ASIC fabricated in TSMC 65\,nm GP.
The ADC array occupies 1.79\,mm$^2$, the PPAC-based spatial equalizer  2.41\,mm$^2$, and the BEACHES engine 0.94\,mm$^2$.
The ASIC also includes an SRAM to perform high-speed tests.
At nominal 1.0\,V supply and a temperature of 300\,K, the ASIC achieves a maximum clock frequency of 312\,MHz with the critical path in the spatial equalizer.
This clock frequency corresponds to 156\,MS/s and 312\,MS/s with 6-bit and 3-bit samples, respectively, as well as to a throughput of 9.98\,Gb/s for 16 UEs transmitting 16-QAM with 6-bit ADC samples.
Under these conditions, reducing the equalizer resolution by 1 bit saves an average of 51\,mW in the spatial equalizer, which corresponds to 18\% of the equalizer's power when operating at its maximum resolution of 4 bits.
The ADC array consumes 106\,mW with the PGAs configured to unit gain.
Furthermore, the BEACHES engine is able to process 9.75\,M channel vectors per second at 79\,mW.
\fref{fig:vfs} shows the equalizer and BEACHES energy when performing voltage-frequency scaling.

Table \ref{tab:comp} compares our ASIC to existing massive MU-MIMO equalizers.
Our implementation is the only one including ADCs and supporting programmable resolution in both the ADC array and equalizer.
After technology normalization, our design achieves $6\times$ higher throughput, $2\times$ higher area efficiency, and $2\times$ lower energy per bit when operating at maximum resolution and taking into account the ADC array area and power consumption. 
Thanks to the resolution-adaptivity of our ASIC, we can further reduce its energy or increase its throughput, up to doubling the throughput and tripling the energy efficiency when operating at the lowest-resolution configuration of 3 ADC bits and 1 equalization bit. 
\begin{table}
\setlength{\tabcolsep}{4pt} 
    \caption{Comparison with state-of-the-art massive MIMO equalizers}
    \label{tab:comp}
    \vspace{-0.2cm}
    \centering
    \scalebox{0.84}{
    \begin{tabular}{@{}lccc|ccccc@{}}
        \toprule
        ~ & \multicolumn{3}{c|}{\multirow{2}{*}{This work}} & Tang & \!\!Tang\!\! & Jeon & Wen & Liu \\
        ~ & \multicolumn{3}{c|}{~} & \cite{tang16a} & \cite{tang17a} & \cite{jeon19b} & \cite{wen20a} & \cite{liu20} \\
        \midrule
        Max.\ BS antennas $B$ & 32 & 32 & 32 & 128 & 128 & 256 & 256 & 128 \\
        Max.\ UEs  & 16 & 16 & 16 & 32 & 16 & 32 & 32 & 8 \\
        Max.\ mod.~[QAM] & 16 & 16 & 16 & 256 & 256 & 256 & 256 & 64 \\
        ADC\,\&{}\,CHEST integrated\!\! & \textbf{yes} & \textbf{yes} & \textbf{yes} & no & no & no & no & no \\
        Resolution-adaptive & \textbf{yes} & \textbf{yes} & \textbf{yes} & no & no & no & no & no \\
        ADC, EQ resolution & 6,4 & 6,4 & 3,1 & N/A & N/A &  N/A &  N/A &  N/A \\
        Includes ADC area \& power\!\! & \textbf{yes} & no & \textbf{yes} & no & no &  no & no & no \\
        \midrule
        Technology~[nm] & 65 & 65 & 65 & 40 & 28 & 28 & 40 & 65$^\textit{a}$ \\
        Core supply~[V] & 1.0 & 1.0 & 1.0 & 0.9 & 1.0 & 0.9 & 1.1 & 1.2 \\
        Core area~[$\text{mm}^2$] & 4.20 & 2.41 & 4.20 & 0.58 & 2.0 & 0.37 & 0.73 & 1.62 \\
        Max. frequency [MHz] & 312 & 312 & 312 & 425 & 569 & 400 & 290 & 500 \\
        Power [mW] & 396 & 290 & 246 & 221 & 127 & 151 & 87 & 120 \\
        Throughput [Gb/s] & 9.98 & 9.98 & 20.0 & 2.76 & 1.80 & 0.354 & 1.96 & 1.5 \\
        \midrule
        Throughput$^\textit{b}$ [Gb/s] & \textbf{9.98} & 9.98 & 20.0 & 1.70 & 0.775 & 0.153 & 1.21 & 1.5 \\
        Area eff.$^\textit{b}$ [Gb/s/$\text{mm}^2$] & \textbf{2.38} & 4.15 & 4.76 & 1.11 & 0.072 & 0.076 & 0.626 & 0.926 \\
        Energy$^\textit{b}$ [pJ/b] & \textbf{39.7} & 29.0 & 12.3 & 260 & 380 & 2\,835 & 96.9 & 80.0 \\
        \bottomrule
    \end{tabular}
    } \\[0.12cm]
    {\scriptsize $^\textit{a}$65nm LP, $^\textit{b}$technology normalized to 65nm at nominal core supply}
\end{table}


\section{Conclusions}
We have shown the first resolution-adaptive all-digital spatial equalizer ASIC for mmWave massive MU-MIMO.
Our design includes an ADC array to support 32 I/Q baseband channels, a processing-in-memory-based spatial equalizer, and a channel estimation engine.
Our ASIC measurements demonstrate that resolution-adaptivity enables highest-in-class throughput while being more than 2$\times$ energy and area efficient compared to state-of-the-art equalizer designs, even with the ADC array power consumption and area included.


\end{document}